%% file: Formatting-Instructions-LaTeX-2024.tex
\documentclass[letterpaper]{article} 
\usepackage{aaai24}  
\usepackage{times}  
\usepackage{helvet}  
\usepackage{courier}  
\usepackage[hyphens]{url}  
\usepackage{graphicx} 
\usepackage{natbib}  
\usepackage{caption} 
\usepackage{graphicx}
\usepackage{amsmath}
\usepackage{booktabs}
\usepackage{algorithm}
\usepackage{algpseudocode}
\usepackage{enumitem}
\usepackage{multirow}

\usepackage{newfloat}
\usepackage{listings}
\DeclareCaptionStyle{ruled}{labelfont=normalfont,labelsep=colon,strut=off} 
\floatstyle{ruled}
\newfloat{listing}{tb}{lst}{}
\floatname{listing}{Listing}
\title{\textit{No-Skim}: Towards Efficiency Robustness Evaluation on Skimming-based Language Models}
\author{
    Shengyao Zhang, Mi Zhang, Xudong Pan, Min Yang
}
\affiliations{
   School of Computer Science, Fudan University, China\\
    shengyaozhang21@m.fudan.edu.cn, {xdpan18, mi\_zhang, m\_yang}@fudan.edu.cn
}

\usepackage{bibentry}

\begin{document}

\maketitle

\input{tex/abs}
\input{tex/intro}
\input{tex/related}
\input{tex/formulation}
\input{tex/method}
\input{tex/exp_set}
\input{tex/exp}

\input{tex/con}

\bibliography{acmart}
\input{tex/app}
\end{document}

%% file: tex/abs.tex
\begin{abstract}
To reduce the computation cost and the energy consumption in large language models (LLM), skimming-based acceleration dynamically drops unimportant tokens of the input sequence progressively along layers of the LLM while preserving the tokens of semantic importance. However, our work for the first time reveals the acceleration may be vulnerable to \textit{Denial-of-Service} (DoS) attacks. In this paper, we propose \textit{No-Skim}, a general framework to help the owners of skimming-based LLM to understand and measure the robustness of their acceleration scheme. Specifically, our framework searches minimal and unnoticeable perturbations at character-level and token-level to generate adversarial inputs that sufficiently increase the remaining token ratio, thus increasing the computation cost and energy consumption. We systematically evaluate the vulnerability of the skimming acceleration in various LLM architectures including BERT and RoBERTa on the GLUE benchmark. In the worst case, the perturbation found by \textit{No-Skim} substantially increases the running cost of LLM by over 145\% on average. Moreover, \textit{No-Skim} extends the evaluation framework to various scenarios, making the evaluation conductible with different level of knowledge.
\end{abstract}


%% file: tex/intro.tex
\section{Introduction}
\label{sec:intro}
In Natural Language Processing, Transformer \cite{vaswani2017attention} has facilitated the birth of pre-trained language models, such as BERT \cite{devlin2018bert}, RoBERTa \cite{liu2019roberta} and GPT \cite{radford2018gpt}, which have brought significant improvements to various downstream applications. Despite the success on the effective performances, the computational complexity and model parameter size are massive, thus deploying these models to real-time service platforms and resource limited (i.e., energy, computation and memory resources) edge devices are very challenging.

\begin{figure}
    \centering
    \includegraphics[width=0.45\textwidth]{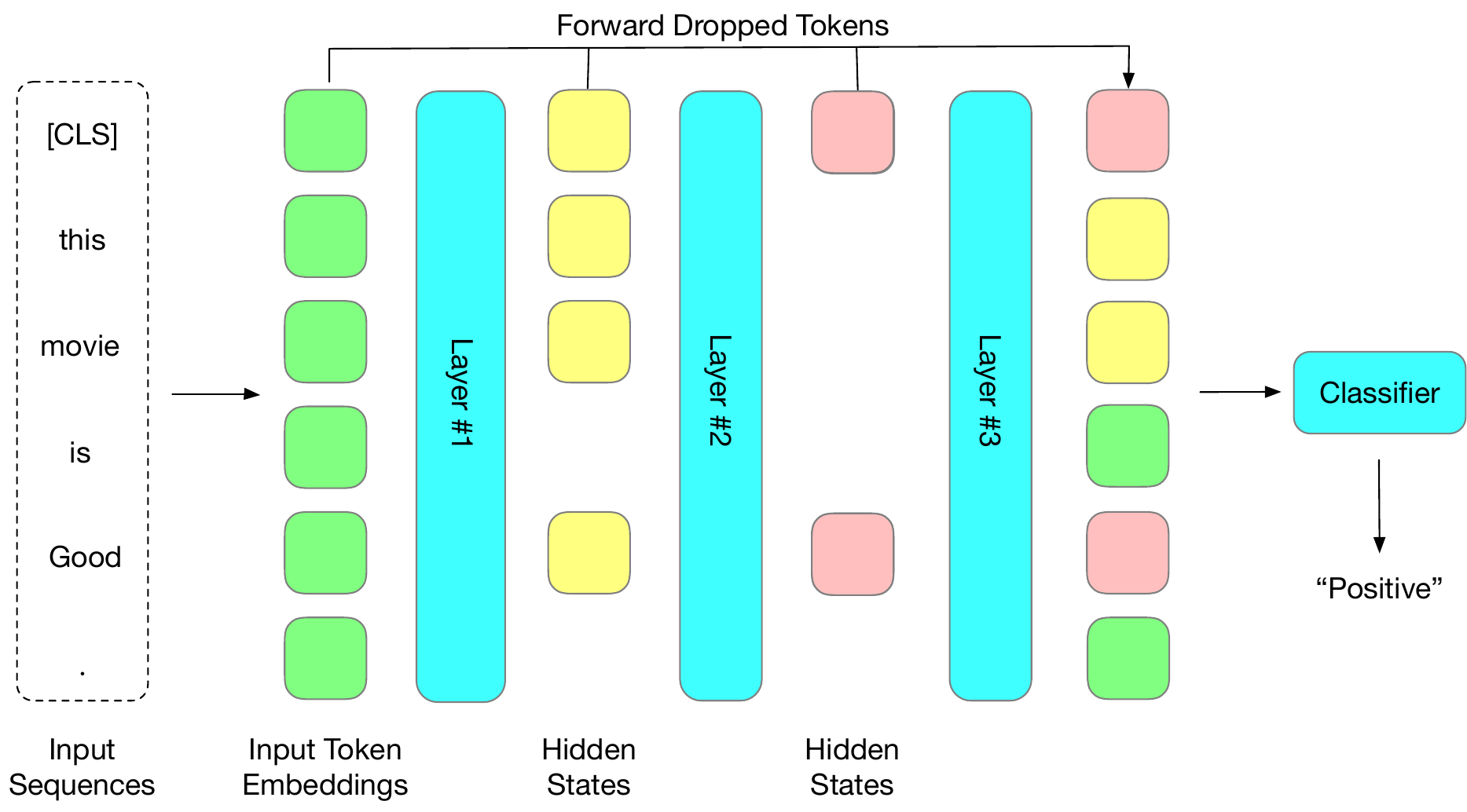}
    \caption{The general design of skimming-based language models on a sentiment classification task. In the example, the important token 
    “good” is preserved.}
    \label{fig:skimming}
\end{figure}

To reduce the computation cost on language models, recent works \cite{goyal2020power,ye2021tr,kim2020length,kim2022learned,guan2022transkimmer} propose the design of skimming-based language models.
Skimming acceleration implements the intuition that human can comprehend the whole sentence by paying extra attention to only a few important words. As shown in Fig. \ref{fig:skimming}, skimming acceleration
dynamically and progressively drops unimportant tokens along different layers to reduce the computation budget and preserves the important tokens within the layers to maintain the semantic information.  For traditional self-attention mechanism, each token in the sequence attends to every other tokens in the input sequence to compute a new feature representation, the computation complexity is quadratic with respect to the input length. Thus, dropping unimportant tokens by skimming can  significantly decrease the computation complexity.

The skimming-based language models can be deployed on real-time service platforms and resource limited edge devices to reduce the computation complexity and decrease the energy consumption. For example, the average FLOPs speed up is 281\% on GLUE benchmark \cite{wang2018glue} when deploying skimming acceleration proposed in \cite{guan2022transkimmer}. 
Despite the tremendous
success on better efficiency, we need to understand and evaluate potential vulnerabilities on existing skimming-based language models from the perspective of computation efficiency, where the decisions of dropping unimportant token can be deliberately manipulated by the unnoticeable but adversarial changes on text inputs. Generating adversarial inputs that may increase the computation complexity will pose serious challenge to the practical deployments. For real-time service platforms, the increasing computation complexity reduces the number of queries processed simultaneously, which eventually damage the service quality. For resource limited edge devices, the increasing computation cost accelerates the consumption of valuable resources (e.g., battery life), which is not acceptable for ordinary users. 

To provide a thorough evaluation on the efficiency robustness of skimming-based language models, we propose \textit{No Skim}, the first general efficiency robustness evaluation framework on the skimming-based language models, which generates adversarial inputs that maximally increase the computation complexity. First, we design the general adversarial inputs generation pipline and divide it into three steps (detailed in Sec. \ref{sec:method}): \textbf{Word Importance Ranking} that identifies the word most relevant to the model efficiency performance, \textbf{Candidate Set Generation} that represents the possible search space of perturbations on original texts and \textbf{Best Candidate Searching} that selects the most effective perturbations on original texts to increase the computation complexity. Second, our evaluation framework is of high modularity, where each step processes the text with different and extensible plug-in components (shown in Fig. \ref{fig:testing}). This ensures the generality of our evaluation framework and makes the evaluation more practical with different levels of knowledge and access to the target skimming-based language model. In details, for word importance ranking, we propose gradient-based and mask-based ranking to accurately identify the important word. For candidate set generation, we propose word-level and character-level perturbations that only require imperceptible changes to maintain the semantic information. For best candidate search, we propose to use direct efficiency indicator (i.e, remaining token ratio) and side-channel information (i.e, inference time) to measure the candidate words' influences on model efficiency.


The contributions of our paper can be summarized as follows:
\begin{itemize}
    \item We are the first work to systematically study the vulnerability of the skimming-based language models from the perspective of efficiency, which poses serious challenges to the practical deployment of skimming-based language models.
    \item We propose an effective and general efficiency robustness evaluation framework \textit{No Skim} that generates adversarial inputs to increase the computation complexity.
    \item We modularize the evaluation framework to be extensible to different plug-in modules. These modules work under various practical scenarios, making the evaluation conductible with three different levels of knowledge.
    \item We conduct extensive evaluations on the state-of-the-art dynamic skimming acceleration scheme Transkimmer \cite{guan2022transkimmer} with BERT and RoBERTa architectures on the GLUE benchmark. In the worst case, our framework can increase the computation cost by 145\%. 
\end{itemize}

%% file: tex/related.tex
\section{Related Works}
\label{sec:related}
\subsection{Skimming Acceleration Schemes}

Skimming acceleration schemes have been a significant thrust of recent research to 
improve the efficiency of existing language models. Skimming is first well-studied in recurrent-based neural networks \cite{yu2017learning,campos2017skip,yu2018fast,fu2018speed}, which saves computation time-wise by dynamically skipping some time steps and copying the hidden states directly to the next step without any update. This matches how human efficiently read texts by emphasizing the important words in the sequence and ignoring parts with little importance. Recently, in the presence of transformer architectures \cite{vaswani2017attention}, skimming-based language models reduce the computation complexity by dropping some unimportant tokens progressively along different layers. Skimming-based acceleration schemes can be categorized into static and dynamic schemes based on whether the remaining token ratio is fix or not. 
\begin{itemize}[leftmargin=*]
\setlength{\itemsep}{0pt}
\setlength{\parsep}{0pt}
\setlength{\parskip}{0pt}
    \item \textbf{Static Skimming Schemes: } Static skimming schemes use a fix remaining token ratio, where all the input sequences are all dropped certain ratio of tokens during inference. \citet{goyal2020power} and \citet{kim2020length} propose PoWER-BERT and LAT respectively, which optimizes a fix remaining token ratio during training. However, different input sequences vary greatly within tasks and between training and validation dataset, leading to a bad generalization.

    \item \textbf{Dynamic Skimming Schemes: } Dynamic skimming schemes are input-adaptive, which use hidden values or attention values to dynamically decide whether the token are dropped or not. \citet{ye2021tr} propose a RL-based scheme called TR-BERT, which adopts reinforcement learning to independently optimize a policy network that dynamically drops tokens. \citet{kim2022learned} propose a threshold-base scheme called LTP. It learns a threshold during training and drops the tokens whose the attention values is lower than the threshold. \citet{guan2022transkimmer} propose a prediction-based scheme called Transkimmer, which integrates each layer with a lightweight fully connected network to make the skimming decision for each token, which predicts a 0/1 mask from the hidden value. In this paper, we mainly investigate the efficiency robustness of the dynamic skimming-based language models as the computation complexity varies according to the input text sequences.
\end{itemize}




\subsection{Efficiency Robustness}
A line of works have been proposed to study the efficiency robustness of existing models. \citet{hong2020sloth} and \citet{zhang-etal-2023-slowbert} propose slow-down attacks on multi-exit models \cite{xin2020deebert,liu2020fastbert,zhou2020patience,kaya2019sdn,huang2017msd} in vision and text domain respectively, which delay the exit positions to increase the computation cost.  \citet{haque2020ilfo} propose ILFO to show the efficiency vulnerabilities of layer skipping networks \cite{wang2018skipnet,figurnov2017spatially}. \citet{haque2022ereba} propose EREBA to increase the models' energy consumption where no the model internal information is accessible. For auto-regression generative tasks, \citet{chen2022nicgslowdown,chen2022nmtsloth} propose to maximize the output sequences' length to increase the inference time. As skimming acceleration schemes improve the model efficiency different from the aforementioned models, a systematic evaluation on the efficiency robustness is necessary.


%% file: tex/formulation.tex
\section{Formulation}

\subsection{Evaluation Objectives}

The goal of our efficiency robustness evaluation framework is to generate adversarial inputs that deteriorate the efficiency of the skimming-based language models. Intuitively, the framework
generates unnoticeable perturbations on original inputs to increase the remaining token ratio, which serves as an indicator of model efficiency. Increasing the remaining token ratio denotes the increase of the computation complexity and the energy consumption. We formulate the evaluation objectives as the following optimization:
\begin{equation}
   \arg\max_{\delta} L_{eff}(x+\delta) \quad s.t. \quad Sim(x,x+\delta) \ge \epsilon,
   \label{eq:formulation}
\end{equation}
where $x$ is the original input, $\delta$ is the perturbations added on the input $x$, $L_{eff}$ denotes the efficiency loss, $Sim: \mathcal{X} \times \mathcal{X} \rightarrow (0,1)$ is the similarity function and $\epsilon$ is the similarity threshold. 

Generating adversarial inputs that increase the computation cost poses serious challenges to the practical deployments of skimming-based language models. For real-time service platforms, throughput is the key element to measure the quality of the service. However, for a platform with a certain level of computation power, the increasing computation complexity reduces the number of queries processed simultaneously, which eventually damages the service quality and ruins online users' experience. For resource limited edge devices, the increasing computation cost accelerates the consumption of valuable resources (e.g., battery life for mobile phones), thus shortening the available time, which is not acceptable for ordinary users.

\begin{figure*}
    \centering
    \includegraphics[width=0.8\textwidth]{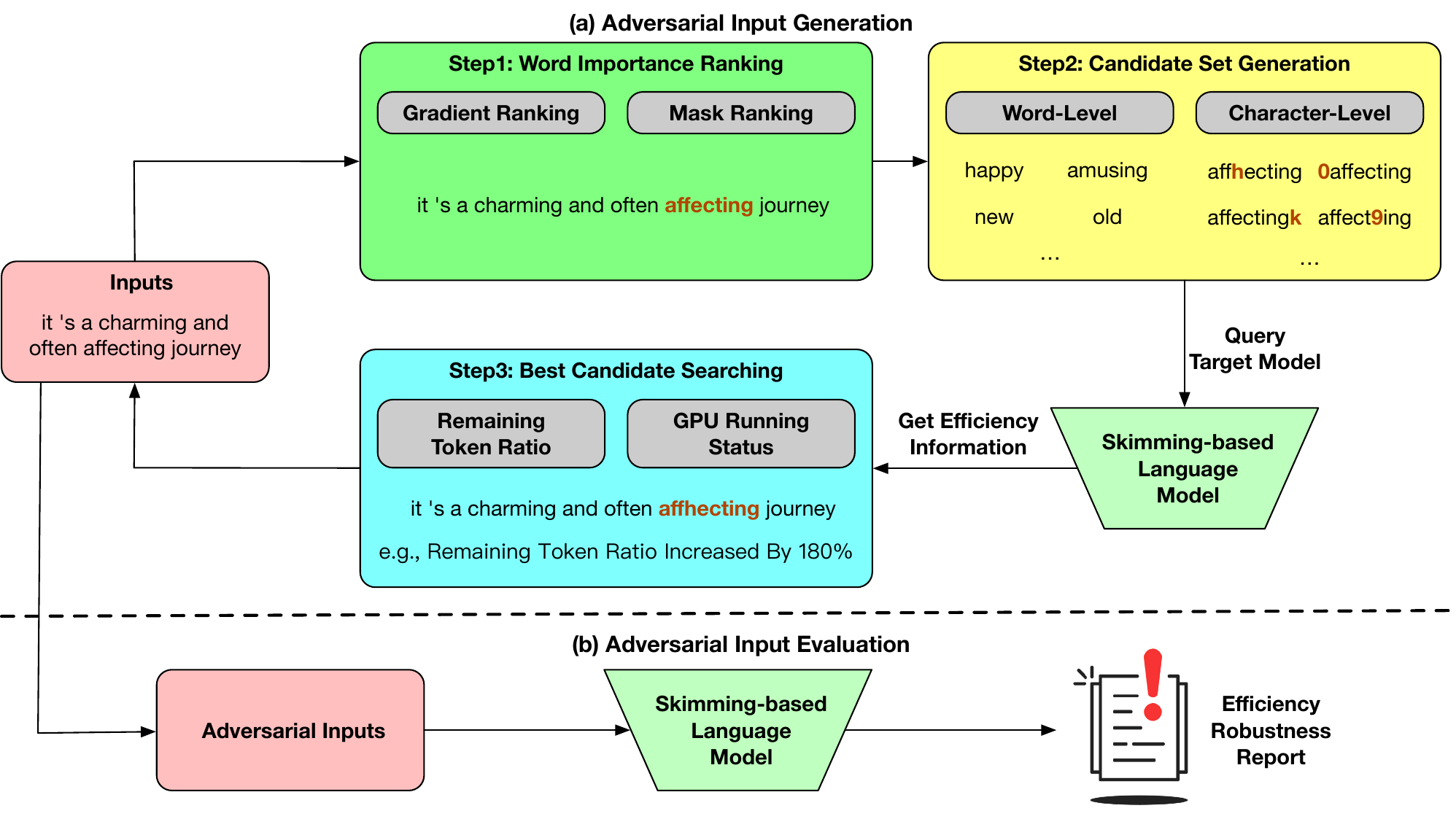}
    \caption{The general evaluation framework of \textit{No Skim}. (a) denotes the adversarial input generation phase, (b) denotes the input evaluation phase. Ranking, generation and searching steps are extensible to other new modules (e.g., shown as boxes with gray backgrounds). }
    \label{fig:testing}
\end{figure*}

\subsection{Evaluation Scenarios}
To comprehensively evaluate the potential efficiency vulnerability of the skimming-based language models, the framework should generally support the evaluation under different level of knowledge and access to the language models. We assume the evaluator has the following knowledge and access to the target skimming-based language model:
\begin{itemize}
    \item \textbf{White-box Access:} White-box access assumes the evaluator has full knowledge of the target model (i.e., model parameters and dynamic skimming scheme) and the vocabulary information (i.e., the vocabulary size and the corresponding word embeddings). And the gradient w.r.t. the word embedding and model parameters can be directly calculated. It simulates a practical scenario where the skimming language model is directly deployed on the resource limited edge devices.
    \item \textbf{Gray-box Access:} Gray-box access assumes that the knowledge of the target model and the vocabulary information is inaccessible to the evaluator. The evaluator only has the ability to obtain the efficiency acceleration information (i.e., the remaining token ratio). It simulates a scenario where the skimming language model is deployed on real-time cloud services to provide efficient prediction and return the speed up information to denote the efficiency.
    \item \textbf{Black-box Access:} Black-box access is considered as the toughest scenario. It assumes that no internal information (i.e., model information or even speed up ratio) are acquirable for the evaluator. The evaluator can only approximate the internal information by observing the running environment status (e.g., GPU running status). It simulates a scenario where the skimming language model is encapsulated as application and then is called by other process on the resource limited edge devices.
\end{itemize}

%% file: tex/method.tex
\section{Methodology}
\label{sec:method}
We propose \textit{No skim}, the first general framework to evaluate the efficiency robustness of the skimming-based language models. First, we present the general design of the framework. Then, we provide three specific implementations to evaluate the efficiency robustness of skimming-based language models under three different scenarios. 

\subsection{General Design}
Given an initial input, our \textit{No skim} iteratively searches unnoticeable perturbations to gradually increase the computation cost. As shown in Fig. \ref{fig:testing}, we provide an overview of the general pipeline of our \textit{No skim}, where Fig. \ref{fig:testing}(a) demonstrates the procedure of each iteration when generating the adversarial input. We also provide a detailed algorithmic description of the generation phase in Algorithm \ref{alg:frame} of Appendix \ref{sec:app-alo}, which consists of three steps:
\begin{enumerate}
    \item \textbf{Step1: Word Importance Ranking (Line 5-8) :} In this step, we aim to identify the most important word that will drastically impact the model's efficiency when modifying it. An importance score will be calculated for each word, and the word with the largest importance score will be modified in the following steps.
    \item \textbf{Step2: Candidate Set Generation (Line 10-12) :} In this step, we aim to generate a candidate set to represent the possible search space given the word selected in the last step, where the candidate words are unnoticeable from the original word to preserve the semantic information.
    \item \textbf{Step3: Best Candidate Searching (Line 14-21) :} In this step, we search the whole candidate set and aim to select the candidate that maximally increases the computation cost and the energy consumption. We substitute the original word with each word in the candidate set and query the target skimming-based language model to get the efficiency information. Then, we compare the efficiency degradation of each candidate word to decide to best candidate word.
\end{enumerate}

As shown in Fig. \ref{fig:testing}(a), it is worth noticing that our general framework modularise the rules of each specific step (i.e., boxes with gray background), which means these three steps are extensible to new modules to further adapt the evaluation framework to more complicated scenarios. 
Fig. \ref{fig:testing}(b) demonstrates the evaluation phase where the generated adversarial inputs are sent to the target skimming-based language model to evaluate the efficiency robustness and generate the final report.

\subsection{White-box Evaluation}

Given a text sequence of n words $X=(x_1,x_2,\cdots,x_n)$, some words play the key role of influencing the model's efficiency. We first calculate the importance score of each word $x_i$ as follows:
\begin{equation}
    Score_{i} = \sum_{j=0}^{m} \frac{\partial{L_{eff(X)}}}{\partial E_{i}^{j}},
    \label{eq:grad_impor}
\end{equation}
where the $E_{i}$ is the embedding of word $x_i$, $m$ is the feature dimension of embedding and $L_{eff}$ is the efficiency loss. The score firstly calculates the gradient of the efficiency loss w.r.t the word embedding and then calculates the sum of gradient along the embedding. The gradient implies the direction and degree of efficiency loss's change when manipulating the word embedding. Perturbing the word with the largest importance score is the easiest way to increase the efficiency loss $L_{eff}$, thus increasing the computation complexity. 

In white-box scenario, we are able to observe the inner characteristic of the target skimming-based model. We directly calculate the remaining token ratio as the efficiency loss:
\begin{equation}
    L_{eff}(X)=  \frac{1}{L} \sum_{l=0}^{L} \frac{sum(M_{l})}{len(M_{l})},
    \label{eq:remaintoeknratio}
\end{equation}
where $L$ is the number of layers in the language model, $M_{l}$ is the binary skim decision for the token sequence at layer $l$. For every element in $M_{l}$, $0$ stands for dropping the token, $1$ stands for preserving the token. The remaining token ratio calculates and averages the ratio of tokens remained in each layer, which represents the computation complexity speed-up.

Once selecting the most important word based on gradient, we need to generate a candidate set composed of unnoticeable perturbed versions of the selected word. The candidate set represents the proper optimization search space to increase the efficiency loss (i.e., remaining token ratio). For white-box scenario, we design word-level perturbation and character-level perturbation to generate the candidate set.

For word-level perturbation, knowing the vocabulary information, we enumerate every word in the vocabulary and calculate the efficiency loss change:
\begin{equation}
    V_{target} = \sum_{j=0}^{m}  (E_{target}^{j}-E_{selected}^{j}) \cdot \frac{\partial{L_{eff(X)}}}{\partial E_{selected}^{j}},
    \label{eq:wordperturb}
\end{equation}
where $E_{target}$ denotes the embedding of a target word in the vocabulary and $E_{selected}$ denotes the embedding of the selected important word. Equation \ref{eq:wordperturb} calculates 
the product of the change of embedding and the partial derivative on embedding when substituting the selected important word to the target word. We then sample words from the vocabulary as follows:
\begin{equation}
    S = \operatorname{Top-k}_{target \in Vocab} V_{target},
\end{equation}
where words with top-k efficiency loss change $V$ are selected. In the meantime, we also discard the words that deteriorate the text semantic information from the candidate set.


For character-level perturbation, we simulate the mistakes made by ordinary users during typing by inserting random characters at random locations. Since the character-level perturbation often leads to UNK token in the embedding space, it is challenging to directly compare the efficiency loss changes of these perturbed words. Thus, we enumerate the characters in digits, letters and insert the character at every possible locations to form the character-level candidate set $S$. The examples of word-level and character-level perturbation are shown in Fig. \ref{fig:testing}(a).

After generating the candidate set, We straightforwardly test all perturbations in the candidate set and select the optimal perturbation that leads to the largest computation cost. In white-box scenario, we use the remaining token ratio calculated in Eq. \ref{eq:remaintoeknratio} to represent the computation cost. For an original input, we iteratively add unnoticeable perturbations to the original input several times to generate highly effective adversarial inputs.

\subsection{Gray-box Evaluation}

In gray-box scenario, the only relative information available is the efficiency acceleration information (i.e., the remaining token ratio in Eq. \ref{eq:remaintoeknratio}). Two types of important information used in white-box scenario is irretrievable.

First, we can not compute the gradient under gray-box setting, which makes the gradient-based importance score in Eq. \ref{eq:grad_impor} impractical. We propose a mask-based importance score to select the word that has the largest impact on the computation efficiency. As proposed in Eq. \ref{eq:gray_score}, we iteratively mask each word $x_i$ in the original text sequence and form the mask version $\hat{X}= (x_1,\cdots,x_{i-1},x_{i+1},\cdots,x_n)$. We then calculate the importance score of each word by subtracting the efficiency loss of the original one from the mask one to get the efficiency loss increment: 
\begin{equation}
    \begin{split}
        Score_{i} &= L_{eff}(\hat{X})-L_{eff}(X),\\
        \operatorname{where}\space\hat{X} &= (x_1,\cdots,x_{i-1},x_{i+1},\cdots,x_n),
    \end{split}
    \label{eq:gray_score}
\end{equation}
where $Score_{i}$ is the important score for the $i'th$ word and $L_{eff}$ represents the remaining token ratio. If masking the word leads to a large efficiency loss increment, it means that the masked word is critical for the model computation efficiency. 

Second, after selecting the candidate word, we can not get the embeddings of words under gray-box scenario. 
In our work, we only generate character-level candidate set to search for the best candidate that maximally increases the computation complexity. The rest of the procedure is the same as the white-box scenario.

\subsection{Black-box Scenario}

\begin{figure}[ht]
    \centering
    \includegraphics[width=0.50\textwidth]{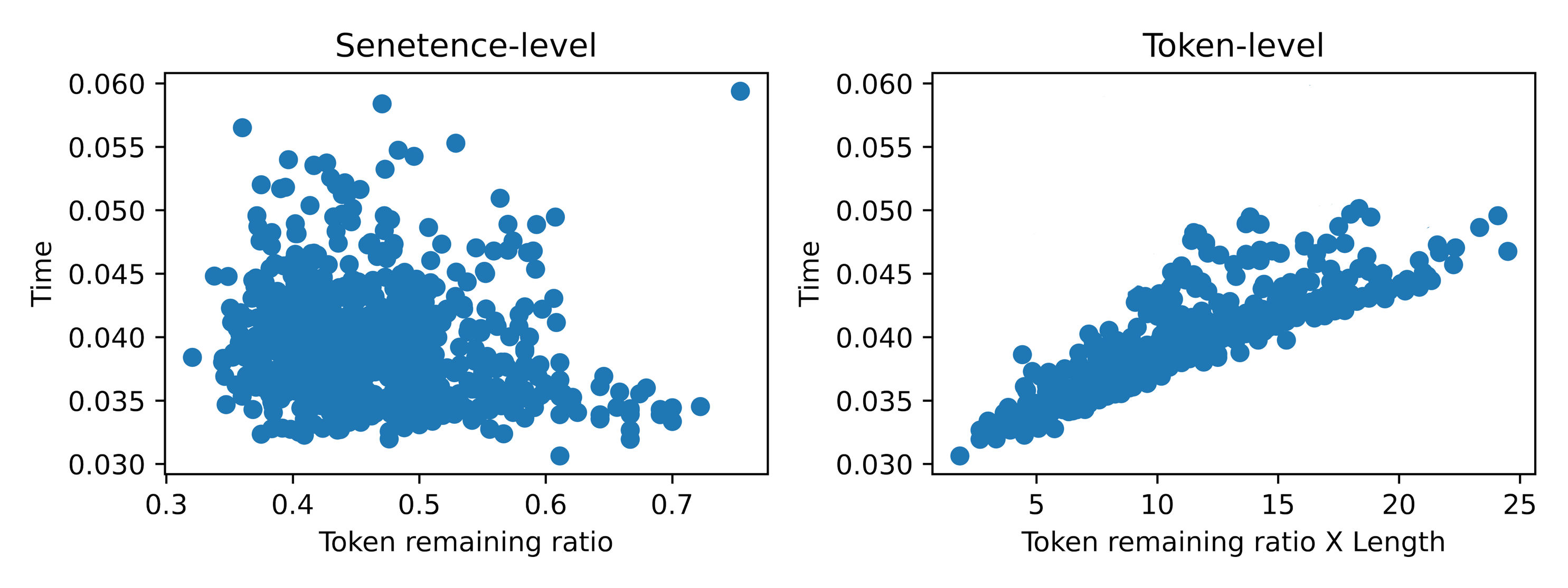}
    \caption{The linear relation between the remaining token ratio and inference time, where (a) uses the inference time on the sequence-level and (b) uses the inference time on the token-level. We present the result on Transkimmer \cite{guan2022transkimmer} with a sentiment classification task SST-2.}
    \label{fig:macs}
\end{figure}


Black-box scenario is considered as the toughest scenario, as no internal information (i.e., model information or even speed up ratio) are acquirable for us. 
Recalled that the goal of skimming-based language models is to reduce the computational complexity by dropping unimportant token during inference phase. Therefore, the computation time during inference inevitably varies with the remaining token ratio. 

Inspired by the recent advances in side-channel attacks \cite{brumley2005time,inci2016cache,goyal2020power,timon2019non}, we propose from a new perspective to approximate the remaining token ratio in Eq. \ref{eq:remaintoeknratio} to facilitate the evaluation procedure, where the magnitude of inference time actually represents different remaining token ratios. First, we empirically observe the relation between the remaining token ratio and the sequence-level inference time, which measures the inference time on the entire input sequence. As shown in Fig. \ref{fig:macs}(a), we find the linear correlation is imperfect due to the large variance of input sequences' lengths. For example, a long input sequence with low remaining token ratio can still require a large inference time.

To get a more precise approximation, we propose token-level inference time to approximate the remaining token ratio and to be set as the efficiency loss $L_{eff}$:
\begin{equation}
    \begin{split}
        L_{eff} &= \frac{Time(x)}{length}, \\
        \operatorname{where} \quad length &= Len(Tokenizer(x)), \\
    \end{split}
    \label{eq:blackbox}
\end{equation}
where $Time(x)$ is the inference of the sequence $x$, $Tokenizer$ is a tokenizer and $Len$ counts the token sequence length. As limited in the black-box scenario, we have no inner information about the architecture of the target model and its corresponding tokenizer. Instead, we propose to randomly select a third-party and public tokenizer unrelated to the targeted skimming-based model to approximate the token sequence length. 

As shown in Fig. \ref{fig:macs}(b), we empirically observe a perfect positive linear correlation between the remaining token ratio and the token-level inference time. The token-level inference time eliminates the negative influence caused by the variance of the sequence lengths, as a single token's computation cost is fixed given a specific language model. This observation proves the feasibility to conduct a side-channel attack to infer the ramaining token ratio by analysis the token-level inference time. After substituting efficiency loss $L_{eff}$ with proposed token-level inference time, the rest of the procedure is the same as the gray-box scenario.

%% file: tex/exp_set.tex
\input{tex/tabs/main}

\section{Evaluation Setting}
\label{sec:eva-set}
\subsection{Architecture \& Dataset}
To thoroughly evaluate our framework, \textit{No skim}, we consider the state-of-the-art skimming scheme Transkimmer \cite{guan2022transkimmer} as our evaluation target. We implement BERT \cite{devlin2018bert} and RoBERTa \cite{liu2019roberta} with tasks of the GLUE Benchmark \cite{wang2018glue}, which are detailed in Tab. \ref{tab:dataset}.


\subsection{Metrics} 
\begin{figure}[ht]
    \centering
    \includegraphics[width=0.48\textwidth]{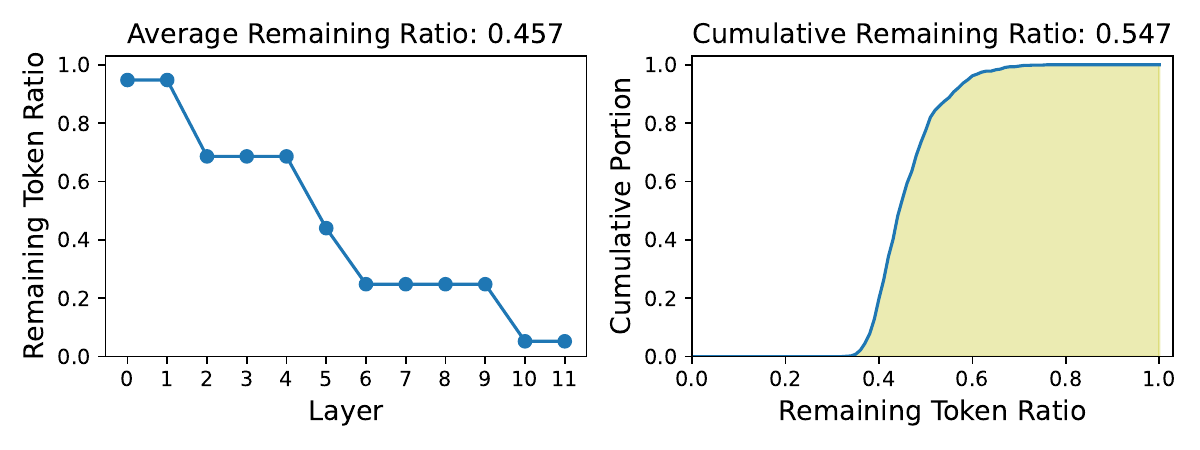}
    \caption{The metrics that evaluate the model efficiency, where (a) denotes the average remaining ratio and (b) denotes the cumulative remaining ratio.}
    \label{fig:metirc}
\end{figure}

We evaluate the efficiency of skimming-based language models with the following metrics: 

\textbf{Average Remaining Ratio (ARR):} As shown in Fig. \ref{fig:metirc}(a), the average token ratio calculates the average remaining token ratio on the dataset. The metric is a task-level metric that evaluate the overall efficiency speed-up performance. When $ARR$ is closer to 0, the target model has better efficiency. 

\textbf{Cumulative Token Ratio (CRR):} 
As shown in Fig. \ref{fig:metirc}(b), the cumulative token ratio calculates cumulative distribution of the remaining token ratio of each input. The metric indicates the variance of efficiency speed-up on different samples. When $CRR$ is closer to 1, the target model has better efficiency. More detailed description are provided in Appendix \ref{sec:app-eval_set}.

\input{tex/tabs/dataset}

\subsection{Detailed Settings.}
We implement the skimming-based language models on the base of Hugging Face’s Transformers  with GLUE benchmark provided in Datasets . 
For training skimming-based models, we fine-tune the pretrained models with a linear classifier. For Transkimmer, we set the skim factor as 0.5 and maximum sequence length as 64. For constructing the adversarial inputs, we select to mutation magnitudes $\epsilon$ from 1 to 5, where each mutant changes one word. For gradient-level mutants, we ensure the semantic similarity between words in the candidate set and original word larger than 0.5. For black-box scenario, we select bert-base-uncased \cite{devlin2018bert} as the third-party and public tokenizer. Since we are the first work to evaluate the efficiency robustness of skimming-based language models, we do not compare with any baseline. Other details are provided in Appendix \ref{sec:app-eval_set}. 
 

%% file: tex/tabs/main.tex
\begin{table*}[!ht]
\begin{center}
\small
\scalebox{0.75}{
\begin{tabular}{cccccccccccccc}
\hline
\textbf{} & \textbf{} & \multicolumn{6}{c}{\textbf{Average Remaining Ratio $\uparrow$}} & \multicolumn{6}{c}{\textbf{Cumulative Remaining Ratio $\downarrow$ }} \\ \cmidrule(l){3-8} \cmidrule(l){9-14} 
\textbf{} & \multicolumn{1}{l}{} & \textbf{Origin} & \textbf{$\epsilon=1$} & \textbf{$\epsilon=2$} & \textbf{$\epsilon=3$} & \textbf{$\epsilon=4$} & \textbf{$\epsilon=5$} & \textbf{Origin} & \textbf{$\epsilon=1$} & \textbf{$\epsilon=2$} & \textbf{$\epsilon=3$} & \textbf{$\epsilon=4$} & \textbf{$\epsilon=5$} \\ \hline
\multirow{4}{*}{\textbf{\begin{tabular}[c]{@{}c@{}}BERT\\ SST-2\end{tabular}}} & \textbf{WhiteBox-Token} & \multirow{4}{*}{0.458} & 0.583 & 0.603 & 0.619 & 0.632 & 0.643 & \multirow{4}{*}{0.547} & 0.422 & 0.402 & 0.387 & 0.373 & 0.362 \\
 & \textbf{WhiteBox-Char} &  & 0.594 & 0.620 & 0.641 & 0.659 & 0.674 &  & 0.411 & 0.385 & 0.364 & 0.346 & 0.332 \\
 & \textbf{GrayBox-Char} &  & 0.508 & 0.543 & 0.571 & 0.592 & 0.610 &  & 0.497 & 0.462 & 0.434 & 0.413 & 0.395 \\
 & \textbf{BlackBox-Char} &  & 0.512 & 0.548 & 0.576 & 0.599 & 0.617 &  & 0.493 & 0.457 & 0.429 & 0.406 & 0.388 \\ \hline
\multirow{4}{*}{\textbf{\begin{tabular}[c]{@{}c@{}}RoBERTa\\ SST-2\end{tabular}}} & \textbf{WhiteBox-Token} & \multirow{4}{*}{0.343} & 0.423 & 0.443 & 0.461 & 0.476 & 0.488 & \multirow{4}{*}{0.662} & 0.582 & 0.562 & 0.544 & 0.529 & 0.517 \\
 & \textbf{WhiteBox-Char} &  & 0.406 & 0.418 & 0.428 & 0.435 & 0.441 &  & 0.599 & 0.587 & 0.577 & 0.570 & 0.564 \\
 & \textbf{GrayBox-Char} &  & 0.404 & 0.409 & 0.412 & 0.413 & 0.414 &  & 0.601 & 0.596 & 0.593 & 0.592 & 0.591 \\
 & \textbf{BlackBox-Char} &  & 0.404 & 0.409 & 0.411 & 0.413 & 0.414 &  & 0.601 & 0.596 & 0.594 & 0.592 & 0.591 \\ \hline
\multirow{4}{*}{\textbf{\begin{tabular}[c]{@{}c@{}}BERT\\ MRPC\end{tabular}}} & \textbf{WhiteBox-Token} & \multirow{4}{*}{0.557} & 0.593 & 0.621 & 0.645 & 0.668 & 0.687 & \multirow{4}{*}{0.449} & 0.412 & 0.384 & 0.360 & 0.337 & 0.318 \\
 & \textbf{WhiteBox-Char} &  & 0.609 & 0.646 & 0.684 & 0.716 & 0.743 &  & 0.396 & 0.358 & 0.321 & 0.290 & 0.262 \\
 & \textbf{GrayBox-Char} &  & 0.603 & 0.641 & 0.675 & 0.704 & 0.729 &  & 0.402 & 0.364 & 0.330 & 0.301 & 0.276 \\
 & \textbf{BlackBox-Char} &  & 0.603 & 0.639 & 0.673 & 0.702 & 0.727 &  & 0.402 & 0.365 & 0.332 & 0.304 & 0.279 \\ \hline
\multirow{4}{*}{\textbf{\begin{tabular}[c]{@{}c@{}}RoBERTa\\ MRPC\end{tabular}}} & \textbf{WhiteBox-Token} & \multirow{4}{*}{0.514} & 0.626 & 0.645 & 0.664 & 0.679 & 0.697 & \multirow{4}{*}{0.491} & 0.379 & 0.360 & 0.341 & 0.326 & 0.308 \\
 & \textbf{WhiteBox-Char} &  & 0.664 & 0.670 & 0.701 & 0.728 & 0.754 &  & 0.369 & 0.335 & 0.304 & 0.277 & 0.251 \\
 & \textbf{GrayBox-Char} &  & 0.603 & 0.624 & 0.646 & 0.665 & 0.682 &  & 0.403 & 0.381 & 0.359 & 0.340 & 0.322 \\
 & \textbf{BlackBox-Char} &  & 0.603 & 0.623 & 0.643 & 0.661 & 0.679 &  & 0.402 & 0.382 & 0.362 & 0.344 & 0.326 \\ \hline
\end{tabular}
}
\end{center}
\caption{The effciency results of our generated adversairal inputs.}
\label{tab:sst-main}
\end{table*}

%% file: tex/tabs/dataset.tex
\begin{table}[!ht]
\begin{center}
\small
\scalebox{0.9}{
\begin{tabular}{ccccc}
\hline
   \textbf{Identifier} & \textbf{Task} &  \textbf{Domain} &  \textbf{Length} & \textbf{Size}   \\ \hline
   \textbf{SST-2} & Sentiment & Movie Reviews &  25 & 67k/0.9k \\
   \textbf{MRPC} & Paraphrase & News & 53  & 3.7k/0.5k  \\

\hline
\end{tabular}
}
\end{center}
\caption{Datasets in our evaluation, where the last column reports the training dataset size and validation dataset size.}
\label{tab:dataset}
\end{table}

%% file: tex/exp.tex
\input{tex/tabs/accuracy}
\section{Evaluation}
\label{sec:eva}

    

\subsection{Effectiveness}
First, we report how much computational complexity is increased by our adversarial inputs in Tab. \ref{tab:sst-main}. We make the following observations: (1) Our \textit{No Skim} demostrates the efficiency vulnerability of the existing skimming-based language model, which increases the average remaining ratio by 145\% and decrease the cumulative remaining ratio by 60\%. (2) Adversarial inputs under white-box scenarios degrades the model efficiency the most, which is predictable as we can acquired gradient information to accurately infer the important tokens. (3) The token-level approximation under black-box scenario can accurately approximate the remaining token ratio that degrade the efficiency similar to gray-box scenario.
(4) RoBERTa is more efficiency robust than BERT on both tasks.

\subsection{Sensitivity}
\begin{figure}[ht]
    \centering
    \includegraphics[width=0.48\textwidth]{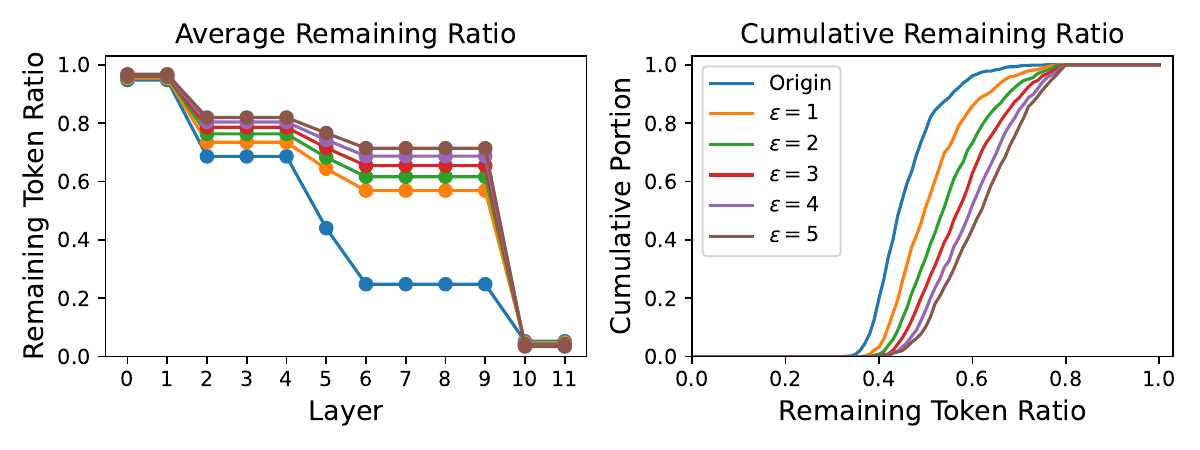}
    \caption{The comparison of efficiency results on BERT and SST-2 using white-box character-level attack.}
    \label{fig:metirc}
\end{figure}

Then, we study how dose the increase of computational complexity react to different magnitude of mutations. As shown in Tab. \ref{tab:sst-main}, with an increased of mutation magnitudes, the corresponding adversarial inputs generated by our \textit{No Skim} effectively degrade the skimming-based models' efficiency to a larger degree. An detailed example of the sensitivity is shown in Fig. \ref{fig:metirc}. In terms of the average remaining ratio, the original inputs efficiently drops the tokens across the layer, while our generated adversarial inputs recover a large portion of the dropped token. And larger magnitude of mutations recover more dropped token. In terms of the cumulative remaining ratio, with larger magnitude of mutations, more inputs have higher remaining token ratio, which conforms to our above observations.

\subsection{Severity}
Next, we study whether our \textit{No Skim} will cause extra damages to the model utility. As reported in Tab. \ref{tab:accuracy}, our attack
generate adversarial inputs that not only increase the
computation costs but also degrade the classification
performances of the skimming-based models. Specifically, the accuracy is decreased from 5\% to 50\%, which calls for more attentions on our proposed evaluation.

\subsection{Stealthiness}
\begin{figure}[ht]
    \centering
    \includegraphics[width=0.48\textwidth]{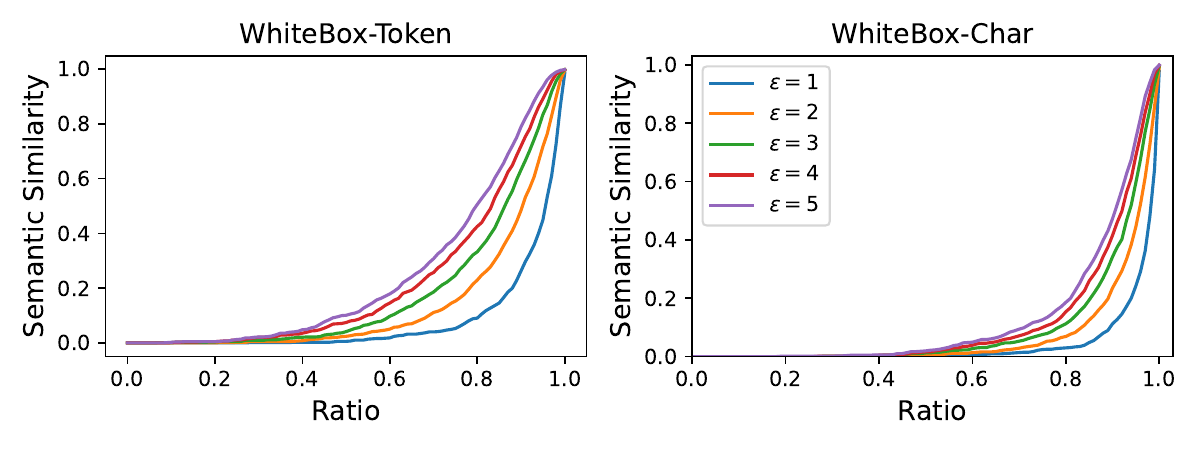}
    \caption{The semantic similarity between adversarial and origin inputs generated by white-box token and character-level perturbations on RoBERTa and SST-2.}
    \label{fig:sim}
\end{figure}

Further, we report how stealthy are the imperceptible mutations on generated adversarial inputs of our \textit{No skim} compared to the original ones. We mainly focus on the cosine simiarity on the sentence embedding generated by SBERT. As shown in Tab. \ref{tab:accuracy}, the average semantic similarity is larger than 0.85 for most cases. The cumulative distribution of the similarity is shown in Fig. \ref{fig:sim}, which also demonstrates that our \textit{No skim} generates highly stealthy adversarial inputs.

\subsection{Rapidness}
Our \textit{No skim} rapidly generate adversarial inputs to evaluate the efficiency robustness of existing skimming-based language models. As shown in Tab. \ref{tab:time} of Appendix \ref{sec:app-res}, most cases only require less than 10 seconds to mutate single word, which is rather acceptable. And the required time increases linearly as the mutation size increases. More analyses are provided in Appendix \ref{sec:app-res}.

%% file: tex/tabs/accuracy.tex
\begin{table*}[!ht]
\begin{center}
\small
\scalebox{0.8}{
\begin{tabular}{ccccccccccccc}
\hline
\textbf{} & \textbf{} & \multicolumn{6}{c}{\textbf{Accuracy (\%)} $\uparrow$} & \multicolumn{5}{c}{\textbf{Semantic Similarity} $\downarrow$} \\ \cmidrule(l){3-8} \cmidrule(l){9-13}
\textbf{} & \multicolumn{1}{l}{} & \textbf{Origin} & \textbf{$\epsilon=1$} & \textbf{$\epsilon=2$} & \textbf{$\epsilon=3$} & \textbf{$\epsilon=4$} & \textbf{$\epsilon=5$} & \textbf{$\epsilon=1$} & \textbf{$\epsilon=2$} & \textbf{$\epsilon=3$} & \textbf{$\epsilon=4$} & \textbf{$\epsilon=5$} \\ \hline
\multirow{4}{*}{\textbf{\begin{tabular}[c]{@{}c@{}}BERT\\ SST-2\end{tabular}}} & \textbf{WhiteBox-Token} & \multirow{4}{*}{0.904} & 0.887 & 0.864 & 0.826 & 0.806 & 0.795 & 0.921 & 0.864 & 0.814 & 0.774 & 0.742 \\
 & \textbf{WhiteBox-Char} &  & 0.889 & 0.852 & 0.823 & 0.777 & 0.753 & 0.920 & 0.842 & 0.842 & 0.693 & 0.693 \\
 & \textbf{GrayBox-Char} &  & 0.869 & 0.848 & 0.799 & 0.759 & 0.729 & 0.916 & 0.843 & 0.775 & 0.716 & 0.668 \\
 & \textbf{BlackBox-Char} &  & 0.881 & 0.847 & 0.812 & 0.774 & 0.741 & 0.920 & 0.845 & 0.782 & 0.723 & 0.673 \\ \hline
\multirow{4}{*}{\textbf{\begin{tabular}[c]{@{}c@{}}RoBERTa\\ SST-2\end{tabular}}} & \textbf{WhiteBox-Token} & \multirow{4}{*}{0.931} & 0.838 & 0.794 & 0.757 & 0.720 & 0.698 & 0.922 & 0.864 & 0.820 & 0.782 & 0.755 \\
 & \textbf{WhiteBox-Char} &  & 0.911 & 0.883 & 0.874 & 0.871 & 0.860 & 0.959 & 0.928 & 0.903 & 0.884 & 0.869 \\
 & \textbf{GrayBox-Char} &  & 0.915 & 0.912 & 0.901 & 0.903 & 0.898 & 0.948 & 0.928 & 0.912 & 0.900 & 0.892 \\
 & \textbf{BlackBox-Char} &  & 0.919 & 0.920 & 0.909 & 0.910 & 0.907 & 0.932 & 0.932 & 0.900 & 0.899 & 0.890 \\ \hline
\multirow{4}{*}{\textbf{\begin{tabular}[c]{@{}c@{}}BERT\\ MRPC\end{tabular}}} & \textbf{WhiteBox-Token} & \multirow{4}{*}{0.853} & 0.767 & 0.696 & 0.642 & 0.586 & 0.556 & 0.951 & 0.927 & 0.907 & 0.888 & 0.872 \\
 & \textbf{WhiteBox-Char} &  & 0.762 & 0.647 & 0.512 & 0.456 & 0.387 & 0.975 & 0.956 & 0.936 & 0.920 & 0.906 \\
 & \textbf{GrayBox-Char} &  & 0.772 & 0.681 & 0.574 & 0.515 & 0.444 & 0.975 & 0.955 & 0.938 & 0.924 & 0.914 \\
 & \textbf{BlackBox-Char} &  & 0.757 & 0.684 & 0.596 & 0.510 & 0.449 & 0.976 & 0.956 & 0.940 & 0.926 & 0.914 \\ \hline
\multirow{4}{*}{\textbf{\begin{tabular}[c]{@{}c@{}}RoBERTa\\ MRPC\end{tabular}}} & \textbf{WhiteBox-Token} & \multirow{4}{*}{0.836} & 0.836 & 0.831 & 0.784 & 0.745 & 0.711 & 0.978 & 0.961 & 0.946 & 0.933 & 0.919 \\
 & \textbf{WhiteBox-Char} &  & 0.802 & 0.735 & 0.721 & 0.662 & 0.632 & 0.973 & 0.954 & 0.936 & 0.924 & 0.910 \\
 & \textbf{GrayBox-Char} &  & 0.831 & 0.792 & 0.757 & 0.730 & 0.689 & 0.977 & 0.960 & 0.942 & 0.924 & 0.910 \\
 & \textbf{BlackBox-Char} &  & 0.851 & 0.826 & 0.772 & 0.748 & 0.735 & 0.976 & 0.960 & 0.942 & 0.925 & 0.911 \\ \hline
\end{tabular}
}
\end{center}
\caption{The negative effect comparison on model utility  and semantic similarity between our generated adversarial inputs and original inputs.}
\label{tab:accuracy}
\end{table*}

%% file: tex/con.tex
\section{Conclusions}
\label{sec:con}

In our work, we systematically study the potential efficiency vulnerability of skimming acceleration schemes on language models. We propose \textit{No Skim}, which generates adversarial inputs that drastically increase the average inference cost of skimming-based language models. This new attack surface poses serious challenges to the deployments of the skimming-based language models on real-time cloud services or local hardware-constrained edge devices. As a security problem of the large language models, our work welcomes future research to devise strong defense against our evaluation.

%% file: tex/app.tex
\appendix
\section{Algorithm}
\label{sec:app-alo}
The detailed algorithmic description of the generation phase is provided below:
\input{tex/alo/general}

\section{Detailed Evaluation Settings}
\label{sec:app-eval_set}

\subsection{Metrics}
We evaluate the efficiency of skimming-based langauge models with the following metrics:

\textbf{Average Remaining Ratio (ARR):}
As shown in Fig. \ref{fig:metirc}(a), the average token ratio first calculates and plots the average remaining token in each layer for all inputs. Then, the metric calculates the normalized area under the curve:
\begin{equation}
    ARR =  \frac{1}{L} \sum_{l=0}^{L} \frac{1}{|D|} \sum_{i=0}^{|D|} \frac{sum(M_{l}^{i})}{len(M_{l}^{i})},
    \label{eq:metric0}
\end{equation}
where $D$ is the test inputs dataset, $L$ is the number of layers and $M_{l}^{i}$ is the binary mask decision at layer $l$ for the $i'th$ test input. The metric is a task-level metric that evaluate the overall efficiency speed-up performance on the entire dataset. When $ARR$ is closer to 0, the target model has better efficiency.

\textbf{Cumulative Token Ratio (CRR):} 
As shown in Fig. \ref{fig:metirc}(b), the cumulative token ratio first calculates and plots the cumulative distribution curve with respect to the remaining token ratio of each input:
\begin{equation}
    p(x_{RTR}) = \frac{1}{|D|}  \sum_{i=0}^{|D|} \{ \frac{1}{L} \sum_{l=0}^{L}  \frac{sum(M_{l}^{i})}{len(M_{l}^{i})} \le x_{RTR} \},
    \label{eq:metirc1}
\end{equation}
where $x_{RTR}$ is the threshold of remaining token ratio and $p(x_{RTR})$ calculates the portion of text inputs that have remaining token ratio larger than $x_{RTR}$. Then, the metric calculates the area size under the curve:
\begin{equation}
    CRR = \int_{0}^{1} p(x_{RTR}) d x_{RTR},
    \label{eq:metric2}
\end{equation}
which is the integral of the portion $p(x_{RTR})$ with respect to the remaining token ratio $x_{RTR}$ on an interval $[0, 1]$. The metric is a sample-level metric that shows the distribution of each sample's remaining token ratio, indicating the variance of efficiency speed-up on different samples. When $CRR$ is closer to 1, the target model has better efficiency.

\subsection{Hyper-parameters}
For training the skimming-based language models, we download
the pre-trained BERT/RoBERTa model provied in Huggingface. and add a linear classifier after [CLS] token embedding. For training Transkimmer on SST-2, we fine-tune the model 3 epochs, where we set the batch size as 32 and the learning rate as 2e-5 with an Adam optimizer. For training Transkimmer on MRPC, we fine-tune the model 5 epochs, where we set the batch size as 32 and the learning rate as 5e-5 with an Adam optimizer. 

\section{More Evaluation Results}
\label{sec:app-res}
\input{tex/tabs/time}
According to Tab. \ref{tab:time}, we make following observations on the time overheads: (1) most cases only require less than 10 seconds to mutate single word, which is rather acceptable. (2) The required time in- creases linearly as the mutation size increases. (3) Gradient-based schemes requires less time than mask-based schemes, as the gradient-based schemes only need one forward/backward process to calculate the word important scores, while mask-based schemes need to multiple forward process to calculate the word importance scores. 

%% file: tex/alo/general.tex
\begin{algorithm}[ht]
\caption{General Efficiency Robustness Evaluations Framework}
\label{alg:frame}

\begin{algorithmic}[1] 
\State{\textbf{Input:} Original Input $X=(x_1,\cdots,x_n)$, Target Skimming-based LLM $F(x)$, Number of Operations $Ops$, Efficiency Loss $L_{eff}(x)$.} 
\State{\textbf{Output:} Adversarial Input Sample $\tilde{X}$.}
\State Initialize: $\tilde{X} \gets X$
\For{range($Ops$)} \Comment{Search perturbations iteratively} 
    \For{word $x_i$ in $\tilde{X}$} 
        \State Compute $Score_{i} \gets \operatorname{ImportanceScore}(x_i)$
    \EndFor 
    \State $X_{sort} \gets \operatorname{Sort}(\tilde{X})$  according to $Score$  \\ \Comment{Step1: Word Importance Ranking} 
    \State $idx \gets \operatorname{Argmax}(Score)$
    \State $x_{max} \gets X_{sort}[0]$
    \State Candidate Set $S \gets \operatorname{CandidateGen}(x_{max})$ \\
    \Comment{Step2: Candidate Set Generation}  
    \State Initialize: $L_{max} \gets L_{eff}(X)$
    \For{candidate word $s_{j}$ in $S$} 
         \State $X_{can} \gets (\tilde{x}_1,\cdots,\tilde{x}_{idx-1},s_{j}, \tilde{x}_{idx+1},\cdots,\tilde{x}_{n})$ 
         \If{$L_{eff}(X_{can}) > L_{max}$}
            \State $s_{best} \gets s_{j}$
            \State $L_{max} \gets L_{eff}(X_{can})$
         \EndIf
    \EndFor \\
    \Comment{Step3: Best Candidate Searching}  
     \State $\tilde{X} \gets (\tilde{x}_1,\cdots,\tilde{x}_{idx-1},s_{best},\tilde{x}_{idx+1},\cdots,\tilde{x}_{n})$
\EndFor
\State \textbf{Return:} $\tilde{X}$
\end{algorithmic}
\end{algorithm}

%% file: tex/tabs/time.tex
\begin{table}[!ht]
\begin{center}
\small
\scalebox{0.8}{
\begin{tabular}{ccccccc}
\hline
\multicolumn{1}{l}{} & \multicolumn{1}{l}{} & \multicolumn{5}{c}{\textbf{Time (sec.)}} \\ \cline{3-7} 
\multicolumn{1}{l}{} & \multicolumn{1}{l}{} & \textbf{$\epsilon=1$} & \textbf{$\epsilon=2$} & \textbf{$\epsilon=3$} & \textbf{$\epsilon=4$} & \textbf{$\epsilon=5$} \\ \hline
\multirow{4}{*}{\textbf{\begin{tabular}[c]{@{}c@{}}BERT\\ SST-2\end{tabular}}} & \textbf{WhiteBox-Token} & 3.124 & 6.977 & 8.555 & 11.434 & 14.275 \\
 & \textbf{WhiteBox-Char} & 4.646 & 21.647 & 32.568 & 37.247 & 42.731 \\
 & \textbf{GrayBox-Char} & 9.031 & 17.744 & 32.551 & 40.598 & 48.704 \\
 & \textbf{BlackBox-Char} & 6.417 & 15.936 & 21.584 & 32.137 & 42.409 \\ \hline
\multirow{4}{*}{\textbf{\begin{tabular}[c]{@{}c@{}}RoBERTa\\ SST-2\end{tabular}}} & \textbf{WhiteBox-Token} & 2.933 & 5.854 & 8.692 & 11.361 & 13.854 \\
 & \textbf{WhiteBox-Char} & 7.974 & 18.043 & 29.157 & 35.711 & 46.077 \\
 & \textbf{GrayBox-Char} & 13.349 & 20.880 & 27.327 & 35.282 & 41.779 \\
 & \textbf{BlackBox-Char} & 14.238 & 22.289 & 29.203 & 37.772 & 44.767 \\ \hline
\multirow{4}{*}{\textbf{\begin{tabular}[c]{@{}c@{}}BERT\\ MRPC\end{tabular}}} & \textbf{WhiteBox-Token} & 3.011 & 6.021 & 8.746 & 11.598 & 14.120 \\
 & \textbf{WhiteBox-Char} & 6.240 & 15.044 & 23.074 & 30.074 & 38.466 \\
 & \textbf{GrayBox-Char} & 10.560 & 21.278 & 30.750 & 42.464 & 52.115 \\
 & \textbf{BlackBox-Char} & 13.529 & 26.282 & 37.214 & 50.571 & 61.181 \\ \hline
\multirow{4}{*}{\textbf{\begin{tabular}[c]{@{}c@{}}RoBERTa\\ MRPC\end{tabular}}} & \textbf{WhiteBox-Token} & 3.165 & 6.528 & 9.431 & 12.370 & 15.430 \\
 & \textbf{WhiteBox-Char} & 8.917 & 15.279 & 21.674 & 30.918 & 37.673 \\
 & \textbf{GrayBox-Char} & 13.279 & 23.541 & 38.498 & 48.570 & 56.381 \\
 & \textbf{BlackBox-Char} & 14.050 & 24.872 & 40.659 & 51.293 & 59.560 \\ \hline
\end{tabular}
}
\end{center}
\caption{Time overheads when generating adversarial evaluation inputs.}
\label{tab:time}
\end{table}